\documentstyle[twoside,fleqn,espcrc2]{article}
\input epsf.tex

\newcommand{\AmS}{{\protect\the\textfont2
  A\kern-.1667em\lower.5ex\hbox{M}\kern-.125emS}}

\begin{document}

\title{\vbox{\hbox{\rightline{\rm\small NSF-ITP-97-110}}
\hbox{Testing M(atrix) Theory at Two Loops}}}
\author{Katrin Becker\thanks{
Address after Sep. 1, 1997: California Institute of
Technology, Pasadena, CA 91125, USA}\\
Institute for Theoretical Physics \\
University of California \\
Santa Barbara, CA 93106
        }

\begin{abstract}

I discuss the relation between M-theory and M(atrix)-theory
in flat space by considering the effective potential for the
scattering of two groups of D0-branes in both theories. 
An explicit calculation of this potential up to two loop order in
M(atrix)-theory reveals a fascinating agreement.

Lecture given at Strings '97; June 17, 1997.

\end{abstract}

\maketitle

\section{Introduction}

This lecture is based on two papers; 
a first paper that was written with Melanie Becker 
\cite{km} and a second paper that we wrote with our collaborators  
Joe Polchinski and Arkady Tseytlin
\cite {bbpt}. 

M-theory is our most promising candidate of being a quantum theory 
in eleven dimensions that includes gravity. Although we do not have a 
complete answer to the question what M-theory is we understand 
several aspects of it: 
\begin{enumerate}
\item All different kinds of string theories can be obtained as different 
compactifications of M-theory \cite{mth}. 
\item At low energies and large distances M-theory reduces to 
eleven-dimensional supergravity. 
\end{enumerate}
An important step towards understanding what 
M-theory is was made 
last year by Banks, Fischler, Shenker and Susskind \cite{bfss}. 
These authors 
conjectured that M-theory in the infinite momentum frame is described 
by a supersymmetric matrix model. The only dynamical degrees of 
freedom or partons are the D0-branes of Polchinski
\cite{joep}, so that the calculation of any
physical quantity in M-theory can be reduced to a
calculation in the M(atrix)-model quantum mechanics. 
The action describing a 
system of $N$ D0-branes
can be regarded as ten-dimensional super 
Yang-Mills theory dimensionally reduced to 0+1 dimensions
\cite{hal,dhn,tow,kp}. The bosonic part of this Lagrangian is 
\begin{equation}
{\cal L} = \frac{1}{2R}  {\rm Tr}\left(- (D_\tau X^i)^2 +
\frac{1}{2}[X^i,X^j]^2 \right),\
\end{equation}
where $R$ is the radius of the eleventh dimension; the signs are
 appropriate for antihermitian $X$
and $i=1, \dots ,9$ denotes the transverse coordinates.

This quantum mechanical system has a $U(N)$ symmetry. While in the 
original formulation of the conjecture, which relates M-theory to 
M(atrix) theory, the $N\rightarrow \infty$ limit was implicit, 
a more recent formulation of the conjecture,
due to Susskind \cite{dlcq,lennyt}, is valid 
for finite $N$. Susskind's new conjecture says: ``The discrete 
light cone quantization of M-theory
\cite{dvv} is exactly described 
by the $U(N)$ M(atrix) theory''.
The discrete light cone quantization agrees with the infinite
momentum frame in the limit $N\rightarrow \infty$.

Our goal is to test this conjecture by making a precise comparison
between M-theory and M(atrix)-theory for finite $N$.
We will do so by computing the effective action for the scattering of
two (groups of) D0-branes.
We will see that the correspondence between both theories is correct 
even at two loops!. 

Let me give an overview of my talk:
\begin{enumerate}
\item I will present the form of the super Yang-Mills action in 
0+1 dimensions that governs the D0-brane behavior. 
\item Derivation of the Feynman rules.
\item Calculation of the one-loop effective action
       of M(atrix)-theory.
\item Calculation of the two-loop effective action of M(atrix)-theory.
\item I will show how the comparison with M-theory 
 works precisely and we will see how on the M-theory side we can make 
more predictions that will agree with M(atrix)-theory calculations.
\item Conclusions and Outlook.
\end{enumerate}

\section{Super Yang-Mills Action in 0+1 Dimensions}
To compute the effective action for two D0-branes it will be
convenient to work with the background field method
\cite{oneloop,lima,abbot}.
This is a technique which allows us to fix a gauge and therefore 
do quantum computations without loosing explicit gauge invariance.
The gauge theory action we are interested in can be obtained 
staring with ten-dimensional super Yang-Mills theory dimensionally reduced
to 0+1 dimensions. After gauge fixing the Lagrangian is

\begin{equation}
{\cal L} = {\rm Tr} \left( \frac{1}{2g}F_{\mu\nu}^2-i
{\bar \psi}D{\psi}+\frac{1}{g}({\bar D}^{\mu}A_{\mu})^2 \right) 
+{\cal L}_{\cal G}\, 
\end{equation}
where $F_{\mu \nu}$ is a $U(2)$ field strength
with $\mu, \nu=0, \dots, 9$, $\psi$ is a real
sixteen component spinor and ${\cal L}_{\cal G}$ is the ghost 
lagrangian.
For the gauge fixing term we will be using the background field 
gauge condition
\begin{equation}
{\bar D}^{\mu}A_{\mu}=\partial^{\mu} A_{\mu} +[B^{\mu},A_{\mu}], 
\end{equation}
where $B_{\mu}$ is the background field.

In 0+1 dimensions we will use the following expressions for the 
field strength and the derivative of the fermionic field
\begin{eqnarray}
& F_{0i}        & = \partial_{\tau} X_i +[A,X_i],\ \nonumber\\
& F_{ij}        & = [X_i, X_j], \nonumber\\
& D_{\tau} \psi & =  \partial _{\tau} \psi +[A,\psi],\ \\  
& D_i \psi      & = [X_i, \psi] \nonumber.  
\end{eqnarray}
Here $A$ denotes the zero component of the gauge field appearing 
in (2). Setting $g=2R$ we recover (1).

We would like to expand the action around a classical background
\begin{equation}
X^i=B^i+{\sqrt g} Y^i,
\end{equation}
that describes the motion of two D0-branes on straight lines, where
\begin{equation}
B^1=i{ v \tau\over 2} \sigma^3 
\qquad {\rm and} \qquad B^ 2 =i{b \over 2}  \sigma^3 .
\end{equation}
Here $v$ is the relative velocity of the two
D0-branes, $b$ is the impact parameter and $\sigma^3$ is a Pauli matrix.
Furthermore $B^i=0$ for $i=0$ and $i=3, \dots 9$.
A convenient form of writing the action is in terms of $U(2)$ generators
by decomposing the fields as \cite{kp}
\begin{equation} 
 A  = {i \over 2} \left( A_0 1\!\! 1 +A_a \sigma^a \right) ,
\end{equation} 
and similarly for the fields $X^i$ and $\psi$. The zero components
of this decomposition describe the motion of the center of mass and will
be ignored in the following.
The Lagrangian is now a sum of four terms
\begin{equation}
{\cal L}={\cal L}_Y+{\cal L}_A+{\cal L}_{\cal G}+{\cal L}_{\rm fermi}, 
\end{equation}
whose explicit form can be found in \cite{km}. Here I only would like 
to mention that the bosonic lagrangians 
${\cal L}_Y$ and ${\cal L}_A$ are described in terms of
sixteen bosons with mass $m_{\cal B}^2=r^2=b^2+(v\tau)^2$, two bosons with
$m_{\cal B}^2=r^2+2v$, two bosons with $m_{\cal B}^2=r^2-2v$ and ten massless 
bosons.
All these fields are real. The ghost action is described in terms of 
two complex bosons with mass $m_{\cal G}^2=r^2$ and one complex massless 
boson.

\section{Feynman Rules}
There are two possible approaches to compute the gauge invariant 
background field effective action.
The first one treats the background field exactly, so that this field
enters in the propagators and vertices of the theory.
To compute the effective action one has to sum over all 1PI graphs
without external lines.
The second approach treats the background field perturbatively, so that 
it appears as external lines in the 1PI graphs of the theory.
We are following the first approach in which we treat the background field
exactly.

We can now proceed to derive 
the Feynman rules.
The explicit form of the vertices can be read off from the actions 
described a moment ago \cite{km}.
The concrete form of the propagators can be easily obtained once we realize
that a relation to the one-dimensional harmonic oscillator can be found.
The propagators of all the bosonic fields take then the form
\begin{eqnarray}
\lefteqn{{\Delta}_{\cal B} \left( \tau,\tau'\vert \mu^2+(v\tau)^2 \right) 
=}\nonumber \\
& & 
\int_0^{\infty} ds e^{-\mu^2 s} \sqrt{
 {v \over 2 \pi \sinh 2sv}}\\
& &\exp \left( -{ v\over 2} 
 \left( {( \tau^2 +\tau'^2 ) \cosh 2sv -2 \tau \tau' )\over \sinh 2sv }
\right)   
\right) \nonumber  ,   
\end{eqnarray}
where $\mu^2=b^2, b^2\pm 2v$ depending on the type of boson that one is 
considering.
The propagator of the fermionic fields is the solution to the
equation
\begin{equation}
\left( -\partial_{\tau}+ m_{\cal F} \right) 
\Delta_{\cal F} \left(\tau,\tau' \vert \ m_{\cal  F}\right) 
=\delta(\tau-\tau'), 
\end{equation}
where $m_{\cal F}=v\tau{\gamma}_1+b {\gamma}_2$ is the fermionic mass matrix.
Using the gamma matrix algebra it is easy to see that the fermionic
propagator can be expressed through the bosonic propagator

\begin{equation}
\Delta_{\cal F} ( m_{ \cal F})=\left( 
\partial_{\tau} +m_{\cal F} \right) 
\Delta_{\cal B} \left(r^2-v\gamma_1
\right) .
\end{equation} 

This is a Dirac-like operator acting on a bosonic propagator of a
particle with mass $r^2-v{\gamma}_1$. Since we have a closed expression 
for ${\Delta}_{\cal B}$, 
we therefore have a closed expression for ${\Delta}_{\cal F}$.
Diagonalizing the mass matrix we find that our theory 
contains eight real fermions with mass $m_{\cal F}^2=r^2+v$ and eight real 
fermions with $m_{\cal F}^2=r^2-v$. The third component
of $\psi$ is massless.
With this Feynman rules we can now proceed to derive the effective actions.
We will start with the one-loop effective action.

\section{One-Loop Effective Action}

In order to compute the one-loop effective action we are interested
in the phase shift ${\delta}$ of one graviton scattered off a second one
\cite{oneloop,lima}

\begin{equation}
\delta =-\int{ d\tau V(b^2+v^2\tau^2). }
\end{equation}

The phase shift can be obtained from the determinants of the operators
$-{\partial}_{\tau}^2+M^2$ that originate from integrating out the massive 
degrees of freedom at one-loop. The result for the one-loop determinants
is \cite{bfss,oneloop}
\begin{eqnarray} 
& &{\rm det}^4 ( -\partial^2_{\tau} 
+r^2 +v){\rm det} ^4(-\partial^2_\tau +r^2 -v) \nonumber \\
& &{\rm det}^{-1} (-\partial^2_{\tau} +r^2+2v )
{\rm det}^{-1} (-\partial_{\tau}^2+r^2-2v) \nonumber \\
& & {\rm det} ^{-6}  (-\partial_{\tau}^2+r^2) . 
\end{eqnarray}
In a proper time representation of the determinants the phase
shift can be written as 
\begin{equation}
\delta= \int_0^{\infty} {ds \over s} 
{e^{-sb^2} \over \sinh sv} \left(3- 4 \cosh sv + \cosh 2sv  \right) 
\end{equation}

For large impact parameter the integrand can be expanded and one
obtains for  
the leading order of the potential the result
\cite {bfss,bercor}
\begin{equation}
V(r)={15 \over 16} {v^4\over r^7} .
\end{equation}
\hyphenation{Sus-skind}
As argued by Banks, Fischler, Shenker and Susskind \cite{bfss}
this is precisely
the result expected for a single supergraviton exchange in eleven
dimensions. Therefore, the $(0+1)$-dimensional M(atrix)-model seems to
know about the propagation of massless modes in eleven dimensions.

Next we would like to check if the agreement between both theories is
specific to one-loop order or if it holds beyond that.
We will do so by computing explicitly the two-loop effective action.

\section{Two-Loop Effective Action}
The two-loop effective action is given by the sum of all 
diagrams of the form contained in Figure 1.
$$
\vbox{
{\centerline{\epsfxsize=3in \epsfbox{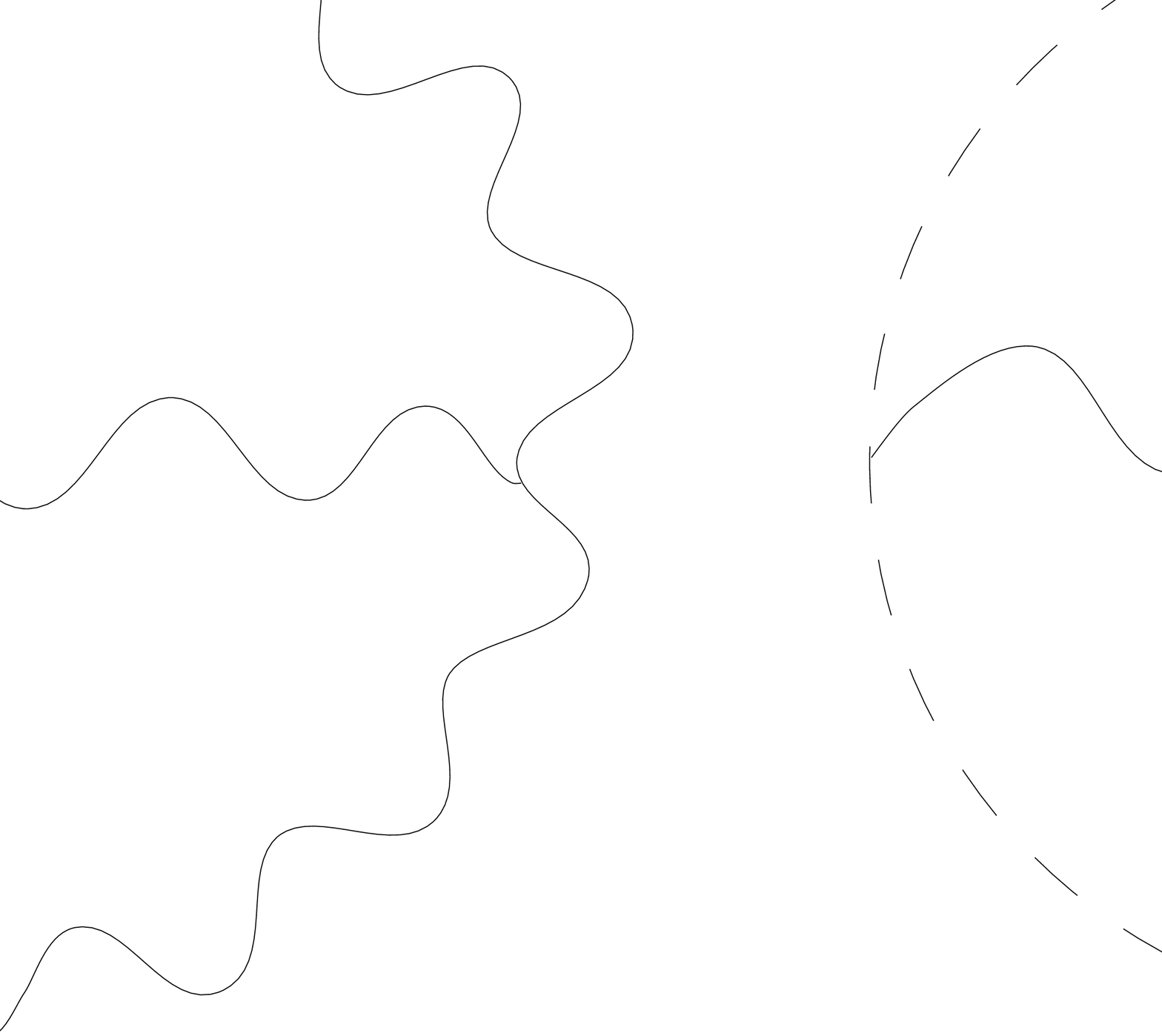}}}
\centerline{Figure 1}
}
$$
The propagators for the fluctuations $Y$
and the gauge field $A$
are indicated by wavy lines, ghost propagators by dashed lines and the 
solid lines indicate the fermion propagators
The explicit expressions for these graphs are
\begin{equation}
\int d \tau {\lambda}_4 {\Delta}_1(\tau,\tau\vert m_1) 
{\Delta}_2(\tau,\tau\vert m_2) , 
\end{equation}
for the diagram involving the quartic vertex ${\lambda}_4$, where 
${\Delta}_1$ and 
${\Delta}_2$ are the propagators of the corresponding particles with 
masses $m_1$ and $m_2$ respectively and
\begin{eqnarray}
\int d \tau d \tau' {\lambda}_3^{(1)} {\lambda}_3 ^{(2)} 
{\Delta}_1(\tau,\tau'\vert m_1) {\Delta}_2
(\tau,\tau'\vert m_2) \nonumber\\
{\Delta}_3(\tau,\tau'\vert m_3)
\end{eqnarray}
for the diagram involving the cubic vertices ${\lambda}_3^{(1)}$ and 
${\lambda}_3^{(2)}$.

Let us see what kind of masses are involved in these diagrams.
In equation (16)  
$m_1$ and $m_2$ are both different from zero.
If, for example, $m_1$ is equal to zero we would have a massless
state running in a loop, that gives a contribution
\begin{equation}
\int {d p \over p^2} ,  
\end{equation}
to the relevant integrals.
However, this expression vanishes in dimensional regularization.
This is because of the following reason.
The dimensional regularization of ill 
defined integrals is defined by three properties \cite{zj}:
\begin{enumerate}
\item translation,
\item dilatation,
\item factorization.
\end{enumerate}
It is the invariance under 
dilatations that imposes the condition that the integral (18) vanishes.
Therefore, diagrams containing a quartic vertex involve two massive
particles. A similar argument for the diagram (17) leads
to the conclusion that exactly one massless state is present.

From dimensional analysis we expect the two-loop effective action to 
be a series of the form
\begin{equation}
\Gamma^{(2)}=g \left( \alpha_0 {1 \over r^2} +
\alpha_2 { v^2 \over r^6} +\alpha_4 {v^4 \over r^{10}} +\dots \right).  
\end{equation}
Odd powers in $v$ in this series are missing because of time reversal
invariance.
The ${\alpha}_i$'s are numerical coefficients that have to be determined 
from the computation of the Feynman diagrams. The details about the
different contributions can be found in \cite {km}.
Here I only would like to mention the final results. 
First, the coefficient of the $v^4/r^{10}$-term appearing at two
loops in matrix theory turns out 
to be equal to zero once all the contributions
coming from bosons and fermions are added up.
The vanishing of this numerical coefficient is in agreement with the
non-renormalization theorem that was conjectured by Banks, Fischler,
Shenker and Susskind \cite {bfss} and it is required in order to 
have agreement with M-theory.

At this moment M(atrix)-theory has passed our first two-loop test:
the vanishing of the $v^4/r^{10}$-term.
However, this is only one term in the effective action of
two D0-branes. By dimensional analysis we observe that the terms 
allowed in the double expansion in $v$ and $r$ take the following form 
\begin{equation}
\begin{array}{cccccccccc}
\displaystyle
{\cal L}_0 &=& c_{00} v^2 &&&&&&& \\
{\cal L}_1 &=&  \displaystyle c_{11} \frac{v^4}{r^7} & +
&\displaystyle c_{12} \frac{v^6}{r^{11}}& + &
\displaystyle c_{13} \frac{v^8}{r^{15}} & +\ \ldots \\[8pt]
{\cal L}_2 &=&  \displaystyle c_{21} \frac{v^4}{r^{10}} & +
&\displaystyle c_{22} \frac{v^6}{r^{14}}& + &
c_{23} \displaystyle\frac{v^8}{r^{18}} & +\ \ldots \\[8pt]
{\cal L}_3 &=&    c_{31} \displaystyle\frac{v^4}{r^{13}} & +
& c_{32} \displaystyle\frac{v^6}{r^{17}}& + &
c_{33} \displaystyle\frac{v^8}{r^{21}} & +\ \ldots
\end{array}
\label{mmseries}
\end{equation}
We have just seen that $c_{12}$ vanishes. Let's see how this result 
appears from the M-theory point of view. 
New surprises for other coefficients in (20) will appear!

\section{Comparison with M-theory and More Predictions}
We consider the scattering of two gravitons with momenta
$p_-=N_1/R$ and $p_-=N_2/R$.
We take $N_1$ to be large so that the first graviton is the source
of the gravitational field. 
The source graviton is taken to have vanishing transverse velocity.
Its worldline is $x^-=x^i=0$ and it produces the Aichelburg-Sexl
metric \cite{aichse}
\begin{equation}
G_{\mu\nu} = \eta_{\mu\nu} + h_{\mu\nu}\ ,
\end{equation}
where the only nonvanishing component of $h_{\mu\nu}$ is
\begin{equation}
h_{--}\ =\ \frac{2 \kappa_{11}^2 p_-}{7 \omega_8 r^7} \delta(x^-)
\ =\ \frac{15 \pi N_1}{ R M^9 r^7} \delta(x^-). \label{hmm}
\end{equation}
Here $\kappa_{11}^2 = 16 \pi^5/ M^{9}$ (see ref.~\cite{bercor} for example),
M is the eleven-dimensional Planck mass up to a convention-dependent
numerical factor and
$\omega_8$ is the volume of the eight-sphere. 
This metric can be thought of as
 obtained from the Schwarzchild metric by
taking the limit of infinite boost in the $+$ direction while the mass is
taken to zero; the latter accounts for the absence of higher-order terms
in
$1/r$ or $N_1$.
The source graviton is in a state of definite $p_-$ and so we
average over the $x^- \in (0,2\pi R)$ direction to give
\begin{equation}
h_{--}\ =\ \frac{15 N_1}{2 R^2 M^9 r^7} \ . \label{hmmav}
\end{equation}

For the action of the `probe' graviton in this field we use the following
trick.  Begin with the action for a massive scalar (spin effects fall off 
more rapidly
with $r$) in eleven dimensions
\begin{eqnarray}
S &=& - m \int d\tau \,(-G_{\mu\nu}\dot x^\mu \dot x^\nu)^{1/2}
\\
&=& - m \int d\tau \, \left( -2\dot x^- - v^2 - h_{--} \dot x^- \dot x^-
\right)^{1/2}\ \nonumber,
\end{eqnarray}
where we have used the form of the Aichelburg-Sexl metric.  A
dot denotes $\partial_\tau $
and $v^2 = \dot x^i \dot x^i$.
This action vanishes if we take $m \to 0$ with
fixed  velocities, but for the process being considered here it is $p_-$
that is to be fixed.  We therefore carry out a Legendre transformation on
$x^-$:
\begin{equation}
p_- = m\, \frac{1 + h_{--} \dot x^-}{\left( -2\dot x^- - v^2 - h_{--}
\dot x^- \dot x^- \right)^{1/2}}\ . \label{legend}
\end{equation}
The appropriate Lagrangian for $x^i$ at fixed $p_-$ is (minus) the
Routhian,
\begin{equation}
{\cal L}'(p_-)\ =\ - {\cal R}(p_-)\ =\ {\cal L} - p_- \dot x^-(p_-) \
.\label{routh}
\end{equation}
Eq.~(\ref{legend}) determines $\dot x^- (p_-)$; it is
convenient before solving to take the limit $m \to 0$, where it reduces
to $G_{\mu\nu}\dot x^\mu \dot x^\nu=0$.  Then
\begin{equation}
\dot x^- = \frac{\sqrt{1 - h_{--} v^2} - 1}{h_{--}}\ .
\end{equation}
In the $m \to 0$ limit at fixed $p_-$ the effective Lagrangian becomes
\begin{eqnarray}
 & {\cal L}' &\to  -p_- \dot x^-
\\
&=& p_- \left\{ \frac{v^2}{2} + \frac{h_{--} v^4}{8} +
\frac{h_{--}^2 v^6 }{16} + \dots \right\}
\nonumber\\
&=& \frac{N_2}{2R} v^2 + \frac{15}{16}\frac{N_1 N_2}{R^3 M^9
}\frac{v^4}{r^7} 
 +   \frac{225}{64 }
\frac{ N_1^2 N_2}{ R^5 M^{18} }\frac{v^6}{r^{14}}  + \dots \nonumber
\label{routhtwo}
\end{eqnarray}

What do we see from this expression
\footnote{The conventions in this section are 
appropriate for hermitian $X$.}?
\begin{enumerate}
 \item The $v$ and $r$ dependences exactly match with the diagonal terms
   of (20) and the $N$-dependence agrees with the leading large-$N$ behavior
    $N^{L+1}$, where $L$ is the number of loops.
\item We see again, that the $v^4/r^7$ term agrees with the 
    one-loop M(atrix)-theory result (15).
\item The absence of a two-loop term $g v^4/r^{10}$ is in agreement 
    with our previous M(atrix)-theory result. 
\item There appears a new term with a coefficient $225/64$ that should
    correspond to a two-loop term in M(atrix)-theory.
\end{enumerate}
Now comes our second two-loop test to M(atrix)-theory. We will ask 
M(atrix)-theory: can you reproduce the two-loop term
with the $225/64$ coefficient for us? Carrying out the calculation of the $v^6/r^{14}$-term in 
M(atrix)-theory by extending the
calculation of \cite{km}, we indeed find precisely the correct
numerical coefficient. The details about the different contributions can be 
found in \cite{bbpt}.
Next we have to reconstruct the $N$-dependence of this result.
Remember, we were considering the scattering of two D0-branes in
M(atrix)-theory. To get the right $N_1$ and $N_2$ dependence, we must
consider the scattering of a group of $N_1$ D0-branes against $N_2$ 
D0-branes.
We can easily reconstruct the $N$-dependence of this scattering process.
In double line notation every graph involves three index loops and so
is of order $N^3$. Terms proportional to $N_1^3$ or $N_2^3$ would only
involve one block (graviton) and so could not depend on $r$.
Symmetry under the interchange of $N_1$ and $N_2$ 
determines that the $SU(2)$ result is multiplied by
\begin{equation}
\frac{N_1 N_2^2 + N_1^2 N_2}{2}\ ,
\end{equation}
which agrees with the supergravity result for the term of interest.
Finally, restoring the dependence on $M$ and $R$ the two-loop result 
of M(atrix)-theory is precisely the result found in the supergravity
calculation (28)!. 

We think that the agreement of both effective actions is fascinating 
since the results for the different contributions to the two-loop
effective action of M(atrix)-theory are rather complicated but
when added up they conspire in such a way that the M-theory result
is exactly reproduced. We are having a hard time believing that this 
agreement is a result of supersymmetry and independent of the M(atrix)-model
conjecture. Therefore, we would like to speculate that there is a different 
structure
behind all this: eleven-dimensional Lorentz-invariance.

\section{Conclusions and Outlook}
M(atrix)-theory has passed several rather strong 
two-loop tests.
First, we have shown that the $v^4/r^{10}$-term at two-loops 
vanishes as it has to be for the agreement with M-theory to be correct.
In our second test we have compared the $v^6/r^{14}$-term 
with the M-theory result. This term in the effective action appearing at two-loops is non-vanishing and exactly agrees in both theories.
This might be indicating that with M(atrix)-theory we are on the right
track.

Some of the many questions that will have to be answered
 in a near future are the following.
In this lecture I have discussed
M(atrix)-theory in flat space. Comparisons of M(atrix)-theory
with supergravity in curved space have been discussed by several authors, see 
for example \cite{dos,ggr,dt}. We think it will be important to find 
the right formulation of M(atrix)-theory in curved space. 
Finally, I would like to remark that for the two-loop calculations
considered so far, the momentum transfer in the eleven-direction has 
been set equal to zero. It will be interesting to understand 
scattering processes of D0-branes
with eleven-dimensional momentum transfer and to make contact 
with the work considered in \cite {keleven}.

\medskip
It is a pleasure to acknowledge the collaboration of Melanie Becker, 
Joe Polchinski and Arkady Tseytlin on the work presented
in this lecture. I would like to thank the organizers of the 
Strings '97 conference for the invitation to present these results.
This research was supported by NSF grant PHY89-04035.

\end{document}